\documentclass{aa} 

\newcommand{\gtsim}{\protect\raisebox{-0.5ex}{$\:\stackrel{\textstyle >}
        {\sim}\:$}}
\newcommand{\ltsim}{\protect\raisebox{-0.5ex}{$\:\stackrel{\textstyle <}
        {\sim}\:$}}

\usepackage{longtable}    
\usepackage{aalongtable}    
\usepackage{supertabular}    
\usepackage{graphicx}
\usepackage{epsfig}
\usepackage{rotating}
\usepackage{lscape}
\usepackage[authoryear]{natbib}
\bibpunct{(}{)}{;}{a}{}{,} 

\begin{document}

\title{Elemental abundances of low-mass stars in the young clusters 25~Ori and $\lambda$~Ori\thanks{Based on {\sc flames} 
observations collected at the Paranal Observatory (Chile). Program 082.D-0796(A).} }
   
\author{K. Biazzo \inst{1} \and S. Randich\inst{1} \and F. Palla\inst{1} \and C. Brice\~no\inst{2}}
\offprints{K. Biazzo}
\mail{kbiazzo@arcetri.astro.it}

\institute{INAF - Osservatorio Astrofisico di Arcetri, Largo E. Fermi 5, 50125 Firenze, Italy 
\and CIDA - Centro de Investigaciones de Astronom\`{\i}a, Apartado Postal 264, M\'erida 5101-A, Venezuela }

\date{Received / accepted }

\abstract
{}
{We aim to derive the chemical pattern of the young clusters 25~Orionis and $\lambda$~Orionis through homogeneous and accurate 
measurements of elemental abundances.}  
{We present {{\sc flames/uves}} observations of a sample of 14 K-type targets in the 25~Ori and $\lambda$~Ori 
clusters; we measure their radial velocities, in order to confirm cluster membership. We derive stellar parameters and 
abundances of Fe, Na, Al, Si, Ca, Ti, and Ni using the code MOOG. } 
{All the 25~Ori stars are confirmed cluster members without evidence of binarity; in $\lambda$~Ori we identify one 
non-member and one candidate single-lined binary star. We find an average 
metallicity [Fe/H]=$-0.05\pm0.05$ for 25~Ori, where the error is the $1\sigma$ standard deviation from the average. 
$\lambda$~Ori members have a mean iron abundance value of $0.01\pm0.01$. The other elements show close-to-solar ratios 
and no star-to-star dispersion. 
}
{Our results, along with previous metallicity determinations in the Orion complex, evidence a small but detectable 
dispersion in the [Fe/H] distribution of the complex. This appears to be compatible with large-scale star formation 
episodes and initial non-uniformity in the pre-cloud medium. We show that, as expected, the abundance distribution
of star forming regions is consistent with the chemical pattern of the Galactic thin disk. 
}
   
	   
\titlerunning{Elemental abundances of 25~Ori and $\lambda$~Ori}
\authorrunning{K. Biazzo et al.}
\maketitle

\section{Introduction}
In the last ten years, several studies have focused on the determination of the chemical composition of open clusters with ages 
$\gtsim 100$ Myr; less attention has been paid to abundances of $\sim 10-100$ Myr old clusters (\citealt{dorazirandich09}, and 
references therein), and to the chemical pattern of young ($\ltsim 10$ Myr) clusters and star forming regions (hereafter, SFRs; 
\citealt{Jamesetal2006, santosetal2008, biazzoetal2011}, and references therein). Instead, elemental abundances of very young clusters 
and SFRs provide a powerful tool to investigate the possible common origin of different subgroups, to investigate scenarios of triggered 
star formation, and to trace the chemical composition of the solar neighborhood and Galactic thin disk at recent times.

Also, whilst the correlation between stellar metallicity and the presence of giant planets around old solar-type stars is well 
established (\citealt{Gonzalez1998, Santosetal2001, johnsonetal2010}) and seems to be primordial (\citealt{Gillietal2006}), 
the presence of a metallicity-planet connection at early stages of planet formation is still a matter of debate. On the one 
hand, the efficiency of dispersal of circumstellar (or proto-planetary) disks, the planet birthplace, is predicted to depend on 
metallicity (\citealt{ercolanoclarke2010}). In a recent study, \cite{yasuietal2010} find that the disk fraction in significantly 
low-metallicity clusters ([O/H]$\sim -0.7$) declines rapidly in $<1$~Myr, which is much faster than the value of $\sim 5-7$ Myr 
observed in solar-metallicity clusters. They suggest that, since the shorter disk lifetime reduces the time available for planet formation, 
this could be one of the reasons for the strong planet-metallicity correlation. On the other hand, \cite{Cusanoetal2011} find 
that a significant fraction of the oldest pre-main sequence stars in the SFR Sh 2-284 have preserved their accretion discs/envelopes, despite 
the low-metallicity environment. Moreover, none of the SFRs with available metallicity determination is metal rich, challenging a full 
understanding of the metallicity-planet connection at young ages. Additional 
measurements of [Fe/H] in several star forming regions and young clusters are thus warranted.

In a recent paper, we have presented an homogeneous determination of the abundance pattern in the Orion Nebula Cluster (ONC) and the 
OB1b subgroup (\citealt{biazzoetal2011}). We report here an abundance study of two other young groups belonging to the Orion complex; 
namely, the 25~Orionis and $\lambda$~Orionis clusters. Although the two clusters have been extensively observed and analyzed, no abundance 
measurement is available for 25~Ori, while for $\lambda$~Ori the metallicity of only one member has been derived 
(namely, Dolan 24; \citealt{dorazietal09}). The metallicity of these two clusters will allow us to put further constrains on the metal 
distribution in the whole Orion complex; furthermore, the age of 25~Ori ($\sim 7-10$ Myr; \citealt{bricenoetal05}) and $\lambda$~Ori 
($\sim 5-10$ Myr; \citealt{dolanmathieu2002}) is comparable to the maximum lifetime of accretion disks as derived from near-IR studies 
and can be considered indicative of the timescale for the formation of giant planets. For these reasons, measurements of [Fe/H] and 
other elements in such clusters are important and timely.

The 25~Ori group is a concentration of T Tauri stars (TTS) in the Orion OB1a subassociation, roughly surrounding the B2e star 
25~Ori; this stellar aggregate was first recognized by \cite{bricenoetal05} during the course of the CIDA Variability Survey of 
Orion (CVSO) as a distinct feature centered at $\alpha_{\rm J2000.0}=$5$^{\rm h}$23$^{\rm m}$ and 
$\delta_{\rm J2000.0}=$1$^{\degr}$45$'$. The 25~Orionis group was also reported by \cite{kharchenkoetal2005} as ASCC-16 in their 
list of 109 new open clusters identified through parallaxes, proper motions, and photometric data. \cite{bricenoetal07} determined 
radial velocities of the cluster candidates, finding that the 25~Ori 
group constitutes a distinct kinematic group from the other Orion subassociations. The color-magnitude diagram, the 
lithium equivalent widths, and the fraction of CTTS ($\sim$6\%) are consistent with the 25~Ori group being older than Ori 
OB1b (age $\sim4-5$ Myr; \citealt{bricenoetal07}). Finally, the same authors derive a cluster radius of $\sim 7$ pc 
centered 23.6$'$ southeast of the star 25~Ori, with some indication of extension farther north.

The $\lambda$~Orionis cluster, discovered by \cite{gomezlada1998}, is distributed over an area of 1 square degree around 
the O8 III star $\lambda^1$~Ori (at a distance of 400 pc; \citealt{murdinpenston1977}). As suggested by 
\cite{dolanmathieu1999, dolanmathieu2001}, the star formation process in this region began some 8--10 Myr ago in the central region 
and has since been accelerating, followed by an abrupt decline $\sim$1--2 Myr ago due to a supernova (SN) explosion 
(\citealt{cunhasmith1996}). CO surveys clearly show that the central region of the cluster, composed by eleven OB stars, 
has been largely vacated of molecular gas. A ring of neutral and molecular hydrogen with a diameter of $\sim$9\degr\,surrounds 
an \ion{H}{ii} region, and is likely the consequence of the  sweeping of interstellar gas by the \ion{H}{ii} region 
(see, e.g., \citealt{langetal2000}, and references therein). 

In Sect.~\ref{sec:obs} we describe the target selection, observations, and data reduction. In Sect.~\ref{sec:analysis} 
the radial velocity and elemental abundance measurements are determined, while the discussion and conclusions are presented 
in Sect.~\ref{sec:discussion} and \ref{sec:conclusion}.

\section{Observations and data analysis}
\label{sec:obs}
\subsection{Target selection}
\label{sec:sample}
We selected four K2-K7 members of 25~Ori from the \cite{bricenoetal07} sample, and four additional stars located about 1 degree to 
the north-east (Brice\~no et al. 2011, in preparation). These four targets, candidate pre-main sequence (PMS) stars in the CVSO 
dataset, were included for follow-up observations with the Hectospec and Hectochelle spectrographs at the 6.5m MMT (Arizona), in order 
to investigate the extent of the 25~Ori aggregate. As for $\lambda$~Ori, we selected four stars from the \cite{dolanmathieu1999} sample 
and two candidates from \cite{barradoetal2004}. 

\begin{figure*}        
\begin{center}
\includegraphics[width=13cm,angle=-90]{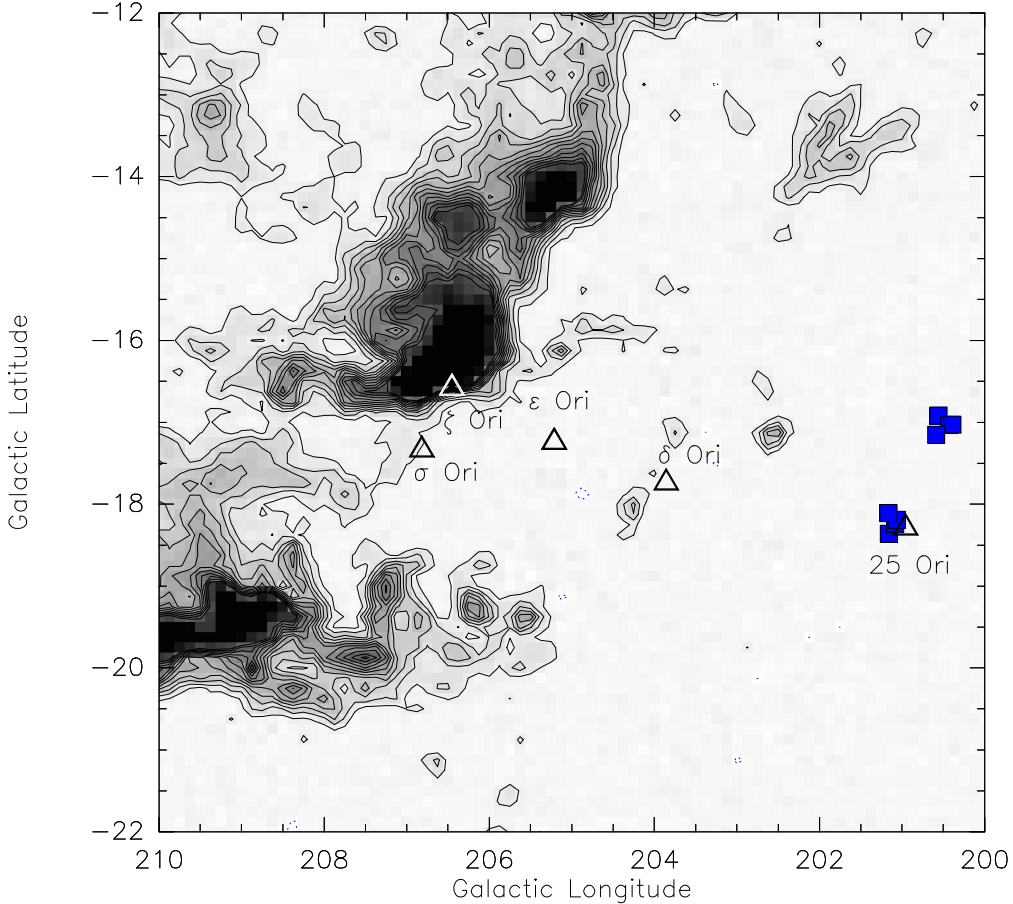}
       \caption{Spatial distribution of the 25~Ori stars (squares) in Galactic coordinates overlaid on the velocity-integrated CO 
       J$=1\rightarrow$0 emission map (\citealt{langetal2000}). The position of 25~Ori, the Orion Belt stars and $\sigma$~Ori is shown 
       by open triangles.}
       \label{fig:25Ori_map} 
 \end{center}
\end{figure*}

\begin{figure*}        
\begin{center}
\includegraphics[width=13cm,angle=-90]{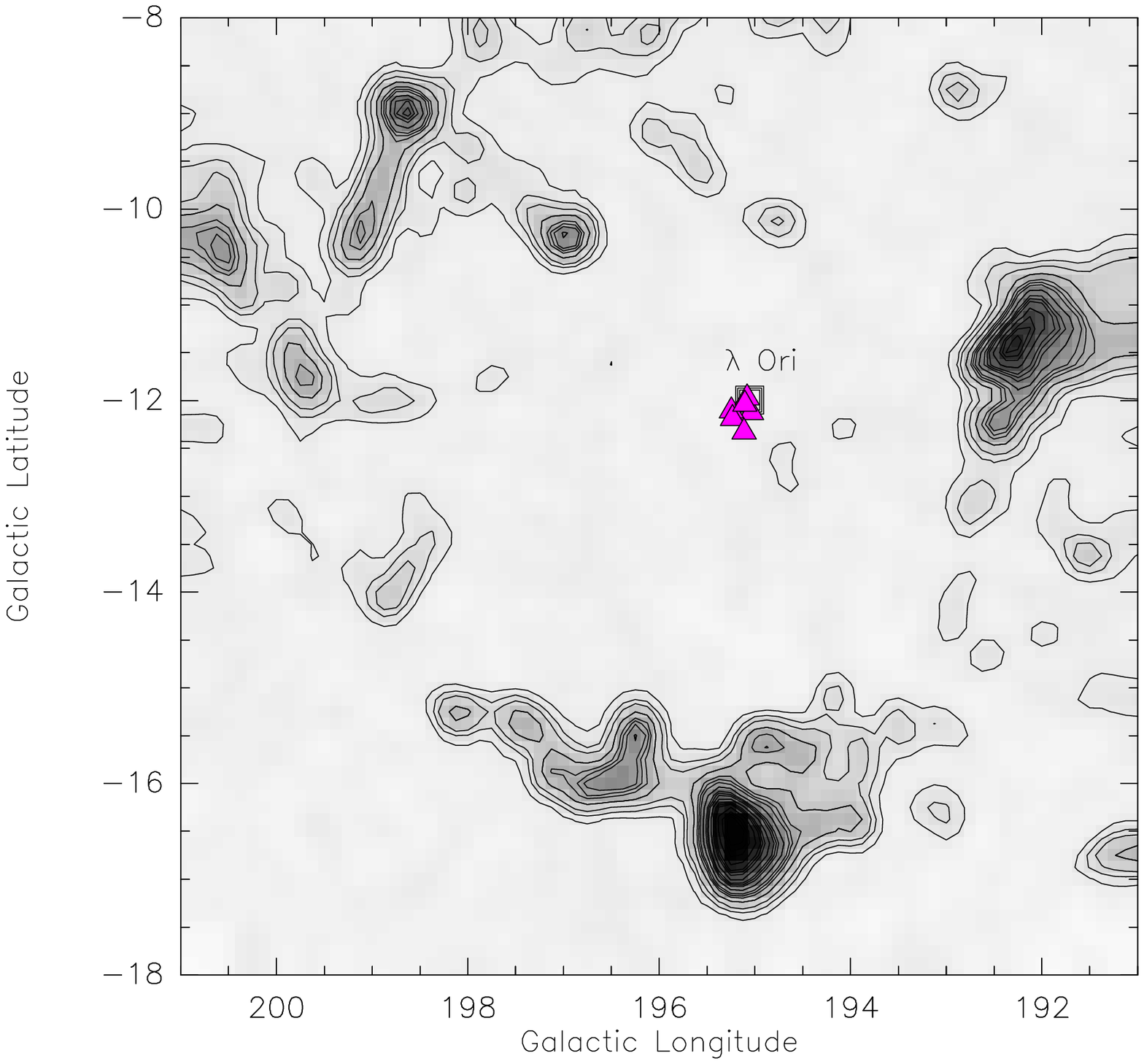}
       \caption{As in Fig.~\ref{fig:25Ori_map}, but for the $\lambda^1$~Ori stars (triangles). The position of $\lambda$~Ori is shown by 
       an open square ($l\sim195$\degr, $b\sim-$12\degr).}
       \label{fig:lambdaOri_map} 
 \end{center}
\end{figure*}

The distribution in the sky of the sample stars is shown in Figs.~\ref{fig:25Ori_map} and \ref{fig:lambdaOri_map} overlaid on CO 
maps\footnote{http://www.cfa.harvard.edu/mmw/MilkyWayinMolClouds.html}. Both samples lie away from giant molecular clouds and 
dark clouds, i.e. far from regions where star formation is actively occurring.

In Table~\ref{tab:literature} we list the sample stars along with information from the literature. In particular, in Columns 1--5 we 
give for the 25~Ori stars the name, $VJK$ magnitude, spectral type, and object class from \cite{Cutri2003}, and Brice\~no et al. 
(2005; 2007; 2011, in preparation), while for the $\lambda$~Ori stars the name, $VJK$ magnitude, 
spectral type, and disk property are from \cite{dolanmathieu1999}, \cite{Cutri2003}, \cite{Zacha2004}, \cite{saccoetal2008}, 
and \cite{hernandezetal2010}. In the last Column of both samples, radial velocities from \cite{dolanmathieu1999}, \cite{bricenoetal07}, 
and \cite{saccoetal2008} are given.

Both samples were selected with the following criteria: $i)$ no evidence of extremely high accretion; $ii)$ no evidence of high rotational 
velocity ($v\sin i \gtsim 30$ km/s), and $iii)$ no evidence of binarity. 

\subsection{Observations and data reduction}
The observations were carried out in October 2008--January 2010 with {\sc flames} (\citealt{pasquinietal02}) on UT2 Telescope. We 
observed both 25~Ori and $\lambda$~Ori using the fiber link to the red arm of {\sc uves}. We allocated four fibers to the 25~Ori 
sample, and two or four fibers to the $\lambda$~Ori sample (depending on the pointing), leaving from two to six fibers for sky 
acquisition. We used the CD\#3 cross-disperser covering the range 4770--6820 \AA~at the resolution $R=47\,000$, allowing us to select 
70+9 \ion{Fe}{i}+\ion{Fe}{ii} lines, as well as 57 lines of other elements (Sect.~\ref{sec:abundance}). 

In the end, the 25~Ori targets were observed in two different pointings, each including four stars, with no overlap. For each 
pointing, we obtained four 45 minutes exposures, resulting in a total integration time of 3 hours each. Two pointings 
were also necessary to observe the $\lambda$~Ori targets. The pointings included four and two stars, and each field was observed 
three times, for a total exposure time of 2.3 hours each. The log book of the observations is given in Table~\ref{tab:observations}.

For a detailed description of the instrumental setup and data reduction, we refer to Biazzo et al. (2011; and references therein). 
Ultimately, the typical signal-to-noise ratio ($S/N$) of the co-added spectra in the region of the Li-$\lambda$6708 \AA~line 
was 30--80 for 25~Ori and 30--150 for $\lambda$~Ori (see Table~\ref{tab:radvel_lithium}). 

\setlength{\tabcolsep}{2.5pt}
\begin{table}[h]  
\caption{Sample stars.}
\label{tab:literature}
\tiny
\begin{center}  
\begin{tabular}{lccclcc}
\hline
\hline
~\\
\multicolumn{7}{c}{25~Ori}\\
\hline
Star$^a$ & $V$ & $J$ & $K$ & Sp.T. & Class$^b$ & $V_{\rm rad}$  \\
 & (mag) & (mag) & (mag)  & &  & (km/s)   \\
\hline
CVSO35  & 14.72  & 11.46 & 10.35 & K7    & C   & 18.8  \\    
CVSO207 & 13.63  & 11.35 & 10.56 & K4    & W   & 24.2  \\    
CVSO211 & 13.98  & 11.78 & 11.06 & K5    & W   & 19.4  \\    
CVSO214 & 13.42  & 11.43 & 10.74 & K2    & W   & 19.1  \\    
CVSO259 & 14.93  & 12.25 & 11.40 & K7.25 & W   & ...   \\  
CVSO260 & ...    & 12.57 & 11.70 & M0.6  & W   & ...   \\   
CVSO261 & ...    & 12.71 & 11.83 & M0    & W   & ...   \\   
CVSO262 & 13.84  & 11.26 & 10.45 & ...   & ... & ...   \\   
\hline
~\\
\end{tabular}
\begin{tabular}{lccclcc}
\multicolumn{7}{c}{$\lambda$~Ori}\\
\hline
Star$^a$ & $V$ & $J$ & $K$ & Sp.T. & Class$^b$ & $V_{\rm rad}$ \\
 & (mag) & (mag)& (mag) &   & & (km/s)  \\
\hline
DM07   & 13.01 & 10.80 & 10.13 & ... &  DL & 25.48 \\
DM14   & 15.18 & 12.07 & 11.19 & K7  &  DL & 25.66, 27.72 \\ 
DM26   & 13.02 & 10.95 & 10.30 & ... &  EV & 24.96 \\
DM21   & 14.80 & 11.66 & 10.84 & ... &  DL & 23.65 \\
CFHT6  & 13.97 & 11.54 & 10.65 & ... &  DL & ...   \\ 
CFHT12 & 14.09 & 11.82 & 10.80 & ... &  DL & ...   \\ 
\hline
\end{tabular}
\end{center}
$^a$ Star names from the following sources: CVSO = CIDA Variability Survey of Orion (\citealt{bricenoetal05}); 
DM = \cite{dolanmathieu1999}; CFHT = Canada-France-Hawaii Telescope 
optical survey (\citealt{barradoetal2004}).\\
$^b$ W: Weak-Lined T Tauri; C: Classical T Tauri; DL: Disk-Less; EV: EVolved disk.
\end{table}  
\normalsize

\setlength{\tabcolsep}{4.pt}
\begin{table}	
\caption{Log of the observations.}
\label{tab:observations}
\begin{center}  
\begin{tabular}{cccccc}
\hline
\hline
$\alpha$    & $\delta$  &  Date      &  UT      & $t_{\rm exp}$ & \#\\
(h:m:s)     &  (\degr: $^\prime$ : \arcsec)  & (d/m/y)    & (h:m:s)  &  (s)          & (stars) \\ 
\hline
~\\
\multicolumn{6}{c}{25~Ori}\\
\hline
05:25:14 & 01:45:32 & 12/10/2008 & 07:17:02 &  2775 & 4\\ 
05:25:14 & 01:45:32 & 16/01/2009 & 02:20:23 &  2775 & 4\\ 
05:25:14 & 01:45:32 & 17/01/2009 & 00:54:27 &  2775 & 4\\ 
05:25:14 & 01:45:32 & 19/01/2009 & 00:37:23 &  2775 & 4\\ 
05:28:06 & 02:51:36 & 21/01/2009 & 00:52:22 &  2775 & 4\\ 
05:28:06 & 02:51:36 & 21/01/2009 & 01:40:26 &  2775 & 4\\ 
05:28:06 & 02:51:36 & 27/11/2008 & 04:58:42 &  2775 & 4\\ 
05:28:06 & 02:51:36 & 29/11/2008 & 06:27:28 &  2775 & 4\\ 
\hline
~\\
\multicolumn{6}{c}{$\lambda$~Ori}\\
\hline
05:34:33 & 09:44:01 &  12/12/2009 & 03:16:07 & 2775 & 4\\
05:34:33 & 09:44:01 &  12/12/2009 & 04:08:18 & 2775 & 4\\
05:34:33 & 09:44:01 &  14/12/2009 & 03:50:03 & 2775 & 4\\
05:35:07 & 09:54:55 &  14/12/2009 & 04:49:03 & 2775 & 2\\
05:35:07 & 09:54:55 &  15/12/2009 & 03:11:49 & 2775 & 2\\
05:35:07 & 09:54:55 &  07/01/2010 & 03:33:58 & 2775 & 2\\
\hline
\end{tabular}
\end{center}
\end{table}  

\begin{table}[h]  
\caption{S/N ratios, radial velocities, luminosities, and masses.}
\label{tab:radvel_lithium}
\tiny
\begin{center}  
\begin{tabular}{lccccc}
\hline
\hline
~\\
\multicolumn{6}{c}{25~Ori}\\
\hline
Star & $S/N$& $V_{\rm rad}$ & Notes$^a$ & $L$&$M^{\rm in}$\\
 & & (km/s) &  & ($L_{\odot}$)& ($M_{\odot}$)\\
\hline
CVSO35  & 70 & 19.8$\pm$1.2 & M & ... & ...\\	 
CVSO207 & 70 & 18.6$\pm$0.6 & M & 0.64& 1.06\\
CVSO211 & 70 & 21.0$\pm$0.4 & M & 0.44& 0.92\\        
CVSO214 & 70 & 20.1$\pm$0.2 & M & 0.62& 1.02\\        
CVSO259 & 50 & 18.9$\pm$0.8 & M & 0.27& 0.74\\
CVSO260 & 30 & 19.2$\pm$0.6 & M & ... & ... \\    
CVSO261 & 30 & 19.8$\pm$0.6 & M & ... & ... \\    
CVSO262 & 80 & 19.2$\pm$1.0 & M & 0.67& 1.07 \\  
\hline
~\\
\end{tabular}
\begin{tabular}{lccccc}
\multicolumn{6}{c}{$\lambda$~Ori}\\
\hline
Star & $S/N$& $V_{\rm rad}$ &  Notes$^a$ & $L$ & $M^{\rm in}$\\
 & & (km/s) & & ($L_{\odot}$)& ($M_{\odot}$) \\
\hline
DM07   & 150 &    25.3$\pm$0.2 &  M & 1.63& 1.35\\
DM14   &  60 &    26.6$\pm$0.4 &  M & 0.46& 0.77\\ 
DM26   & 120 &    28.0$\pm$0.2 &  M & 1.44& 1.22\\
DM21   &  50 &    25.4$\pm$1.2 &  M & 0.67& 0.86\\
CFHT6  &  30 &    26.8$\pm$0.6 &  PB& 0.79& 0.82\\ 
"      &  "  &    31.1$\pm$0.5 &  " & "   & " \\ 
CFHT12 &  60 &  $-$3.3$\pm$2.5 &  NM& ... & ...\\ 
\hline
\end{tabular}
\end{center}
$^a$ M = Member; NM = Non-Member; PB = Probable Binary.
\end{table}  
\normalsize

\section{Analysis and results}
\label{sec:analysis}

\subsection{Radial velocity measurements and membership}
\label{sec:rad_vel}
We measured radial velocities (RVs) in each observing night, to identify possible non-members and binaries. Since we could not 
acquire any template spectrum, we first chose for both 25~Ori and $\lambda$~Ori the stars with the spectra at highest $S/N$, 
low $v\sin i$ ($<$15 km s$^{-1}$), available RV measurements from the literature, and without any accretion signature. In particular, 
we considered CVSO211 for 25~Ori and DM14 for $\lambda$~Ori and followed the method explained in \cite{biazzoetal2011}. 
We obtained $V_{\rm rad}=$21.0$\pm$0.4 km s$^{-1}$ for CVSO211, while for DM14 we found $V_{\rm rad}=$26.6$\pm$0.4 km s$^{-1}$. 
These values are close to the values of $V_{\rm rad}=$19.4$\pm$0.5 km s$^{-1}$ and $V_{\rm rad}=$27.72$\pm$0.18 km s$^{-1}$ 
obtained by \cite{bricenoetal07} and \cite{saccoetal2008}, respectively. Using these two templates, we measured the heliocentric RV 
of all our targets using the task {\sc fxcor} of the 
IRAF\footnote{IRAF is distributed by the National Optical Astronomy Observatory, which is operated by the Association of the 
Universities for Research in Astronomy, inc. (AURA) under cooperative agreement with the National Science Foundation.} package 
{\sc rv}, following the prescriptions given by \cite{biazzoetal2011}. 

All our stars were characterized by single peaks of the cross-correlation function. For all measurements with RV of 
different nights in agreement within $2\sigma$, we computed a mean RV as the weighted average of the different values. 
Only the star CFHT6 in $\lambda$~Ori shows evidence of binarity, with RV measurements around $V_{\rm rad}=$26.8 km s$^{-1}$ 
(in December 2009) and $V_{\rm rad}=$31.1 km s$^{-1}$ (in January 2010). Final radial velocities are listed in 
Table~\ref{tab:radvel_lithium}.

\begin{figure}	
\begin{center}
 \begin{tabular}{c}
  \includegraphics[width=9.2cm]{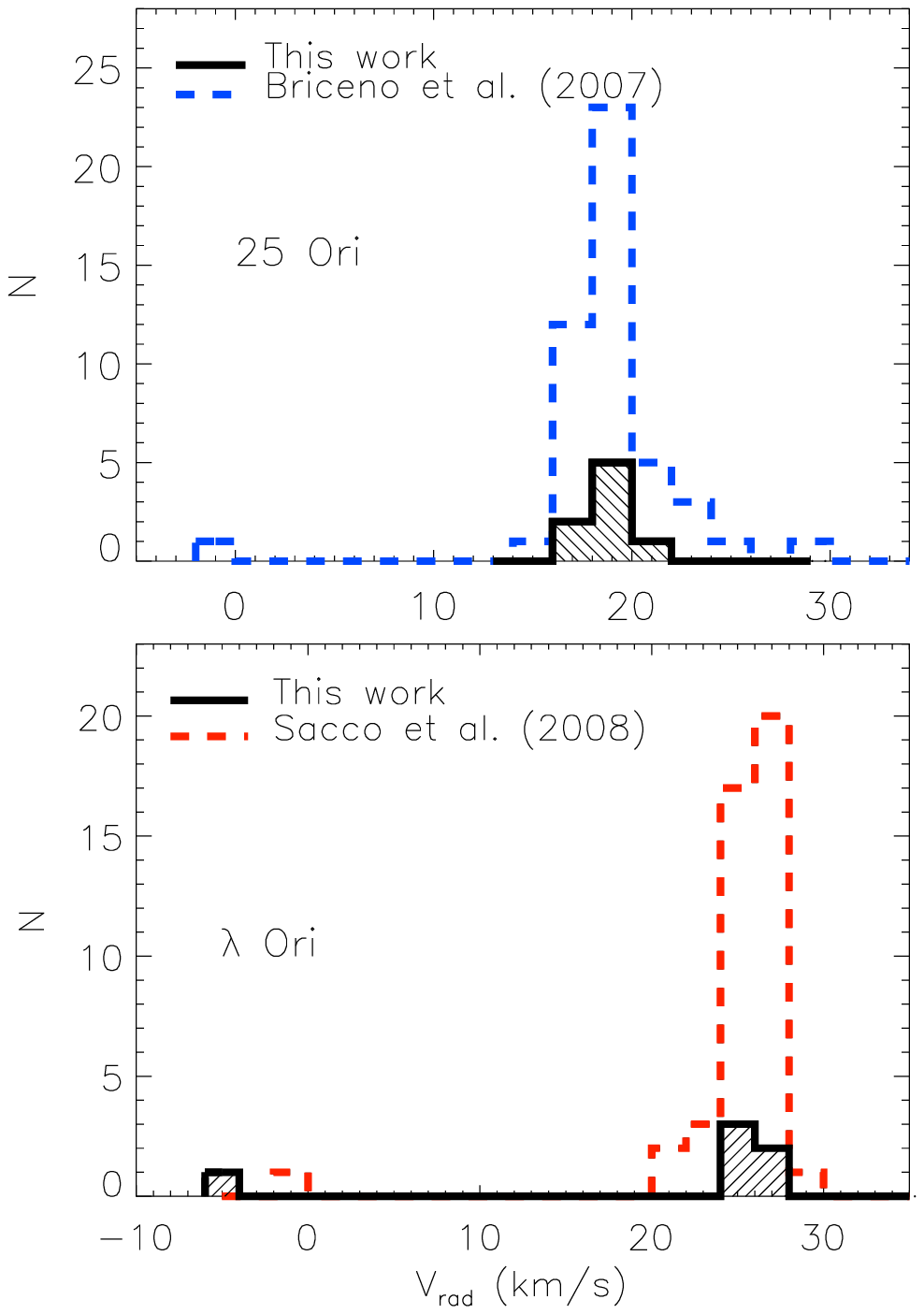}            
 \end{tabular}
       \caption{Radial velocity distributions (solid lines) of all the targets in 25~Ori ({\it upper panel}) and $\lambda$~Ori 
       ({\it lower panel}). The dashed lines represent the distribution obtained by \cite{bricenoetal07} for 47 stars in 25~Ori 
       and by \cite{saccoetal2008} for 45 members in $\lambda$~Ori.}
       \label{fig:vrad_distr}
 \end{center}
\end{figure}

We find that all the 25~Ori stars are members with a cluster distribution centered at $<V_{\rm rad}>=19.6\pm0.8$ km s$^{-1}$ 
(Fig.~\ref{fig:vrad_distr}, Table~\ref{tab:radvel_lithium}), i.e. very close to the peak of $<V_{\rm rad}>=19.7\pm1.7$ km s$^{-1}$ 
found by \cite{bricenoetal07}. In particular, the four new PMS stars off to the north-east of 25~Ori (CVSO259,260,261,262) 
share the same radial velocity of the cluster members from \cite{bricenoetal07}, consistent with being members of this 
stellar aggregate. 
Although four stars represent a small number statistics, this result supports the suggestion made by \cite{bricenoetal07} 
that 25~Ori extends further north than implied by the map in their Fig. 5. The full photometric census of 25~Ori will be
presented in Brice\~no et al. (2011, in preparation).
All the $\lambda$~Ori stars are members, with the exception of CFHT12, as also suggested 
by \cite{barradoetal2004} from photometric selection criteria. Excluding CFHT6 and CFHT12, we find 
a mean $\lambda$~Ori RV of $V_{\rm rad}=$26.3$\pm$1.3 km s$^{-1}$, in good agreement with the cluster distribution 
centered at $<V_{\rm rad}>=27.03\pm0.49$ km s$^{-1}$ (\citealt{saccoetal2008}) . 

\subsection{Final sample for elemental analysis}
\label{sec:final_sample}
\begin{figure*}	
\begin{center}
 \begin{tabular}{c}
  \resizebox{\hsize}{!}{\includegraphics{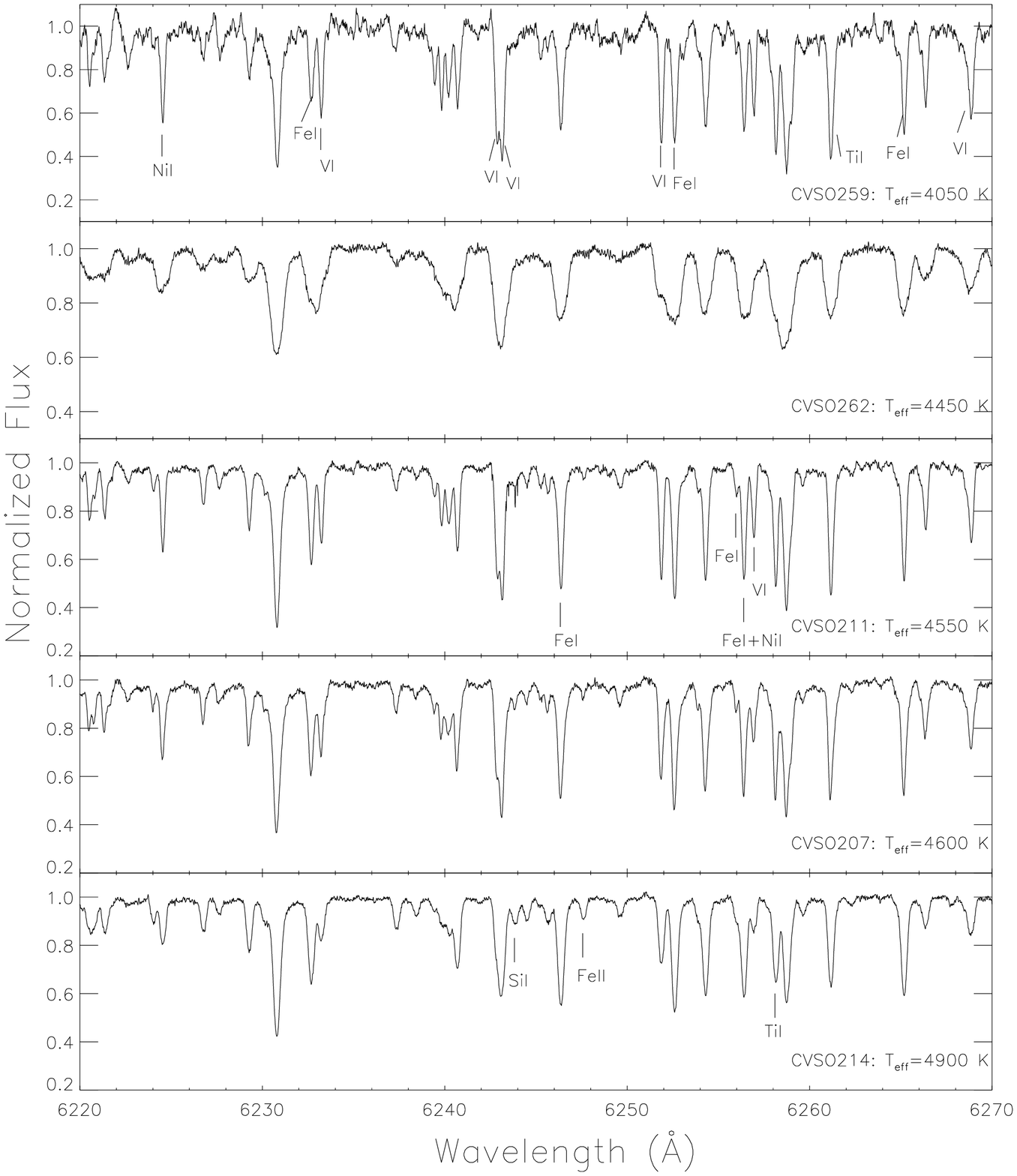}}\\                  
 \end{tabular}
       \caption{Portion of spectra of the targets in 25~Ori used for the abundance determination. The values of $T_{\rm eff}^{\rm SPEC}$ 
       are given for each star.}
       \label{fig:spectra_25}
 \end{center}
\end{figure*}

\begin{figure*}	
\begin{center}
 \begin{tabular}{c}
  \resizebox{\hsize}{!}{\includegraphics{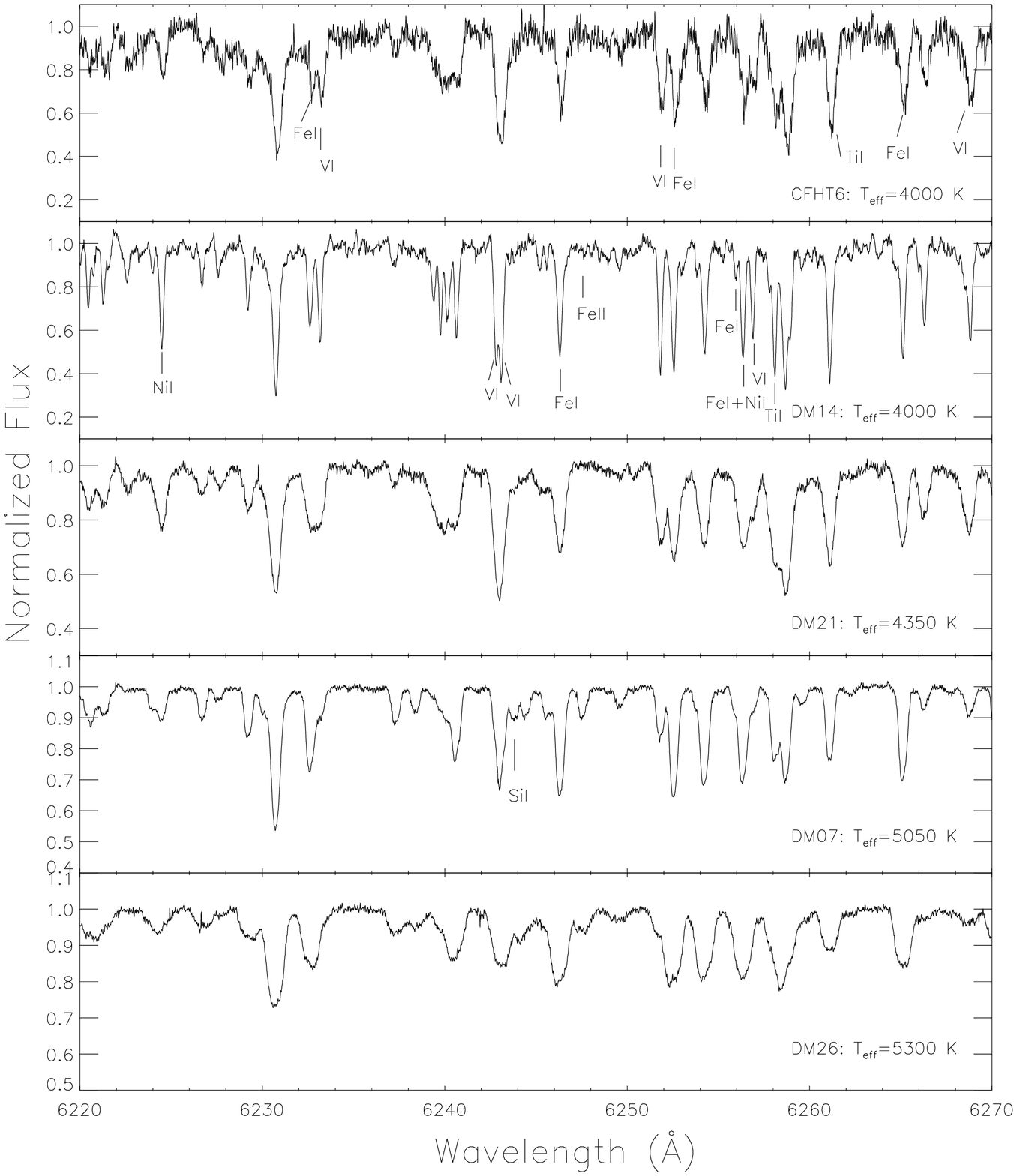}}\\                  
 \end{tabular}
       \caption{As in Fig.~\ref{fig:spectra_25}, but for $\lambda$~Ori. }
       \label{fig:spectra_l}
 \end{center}
\end{figure*}

Unfortunately, some of the stars are not suitable for abundance analysis. In particular, CVSO35 is a very cool accreting 
star (classified as CTTS by \citealt{bricenoetal05}), for which the abundance determination through equivalent widths is not 
reliable. Moreover, CVSO260 and CVSO261 have low $S/N$ spectra ($\ltsim 40$) and low effective temperature 
($T_{\rm eff}\ltsim$4000 K). In these $T_{\rm eff}$ regime, strong molecular bands do not allow one to derive accurate EWs. 
As a result, we performed abundance measurements in ten stars, five in 25~Ori and 
five in $\lambda$~Ori. In Figs.~\ref{fig:spectra_25} and \ref{fig:spectra_l} we display the co-added spectra of these stars 
in the spectral region (6220--6260 \AA) containing several features used to derive the effective temperature through 
line-depth ratios (Sect. \ref{sec:initial_parameters}), and to measure the elemental abundances (Sect. \ref{sec:abundance}).

\subsection{Initial temperature and surface gravity}
\label{sec:initial_parameters}
\begin{figure}	
\begin{center}
 \begin{tabular}{c}
  \includegraphics[width=9cm]{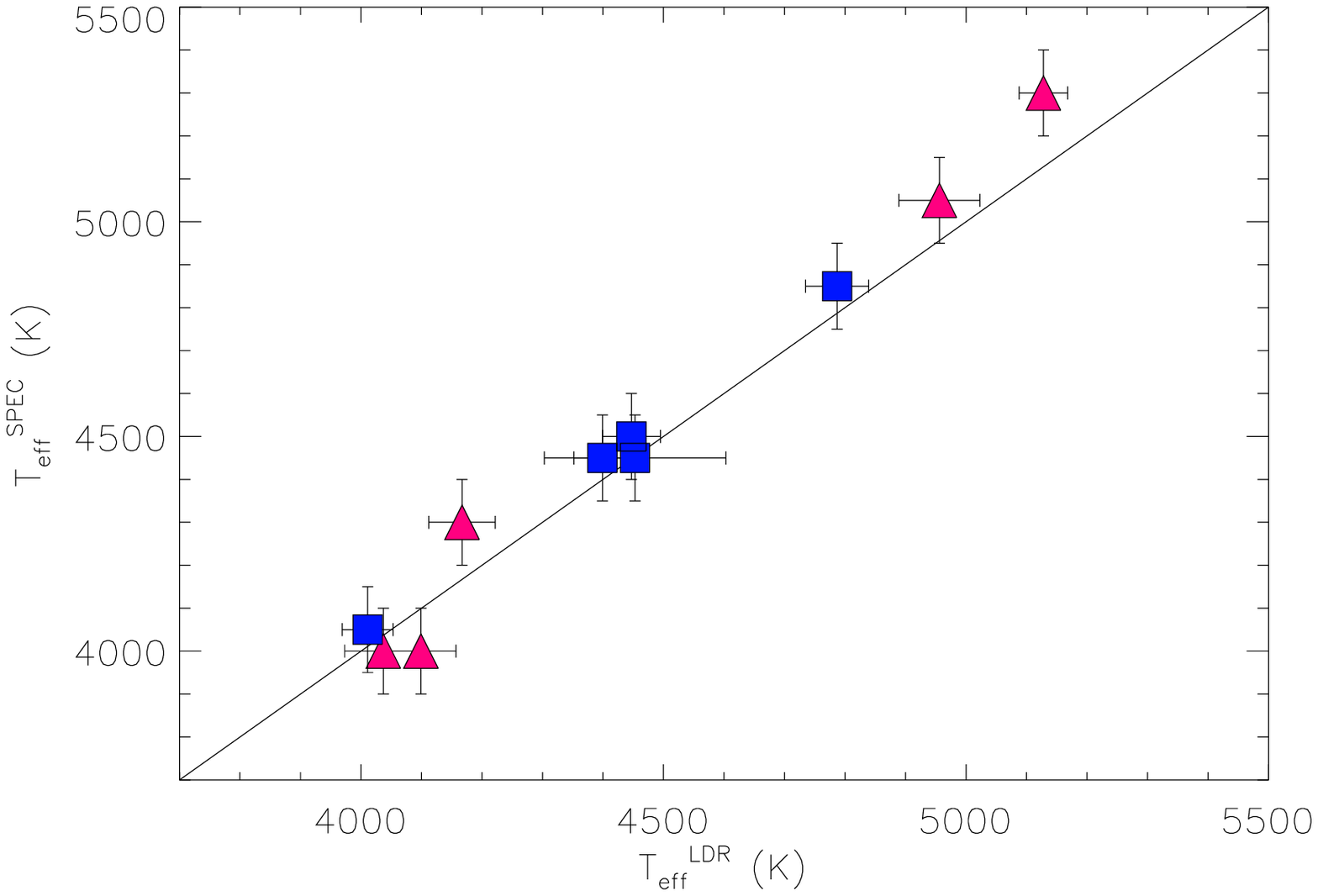}                  
 \end{tabular}
       \caption{Spectroscopic effective temperatures ($T_{\rm eff}^{\rm SPEC}$) derived by imposing the excitation 
       equilibrium versus effective temperatures obtained through line-depth 
       ratios ($T_{\rm eff}^{\rm LDR}$). The squares and triangles refer to 25~Ori and $\lambda$~Ori, respectively.}
       \label{fig:teff_ldr_spec}
 \end{center}
\end{figure}

Since effective temperatures from the literature were not available for all our targets, we used as initial 
the values ($T_{\rm eff}^{\rm LDR}$) derived using the method based on line-depth ratios (\citealt{Gray1991}). 
We followed the prescriptions given by \cite{biazzoetal2011} and obtained the values listed in Table~\ref{tab:param_chem} 
and plotted in Fig.~\ref{fig:teff_ldr_spec}.

\begin{figure}	
\begin{center}
  \includegraphics[width=9cm]{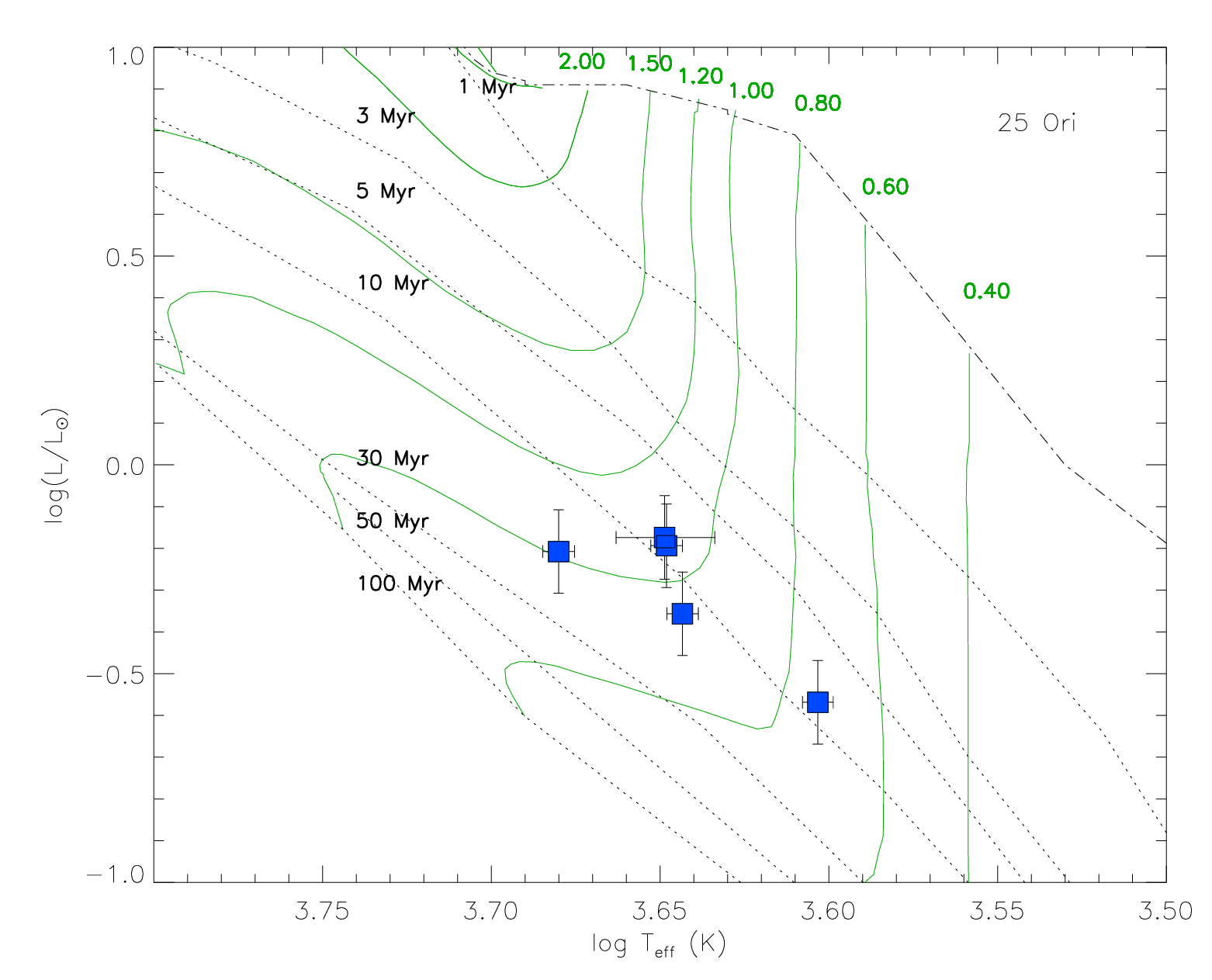}                  
  \includegraphics[width=9cm]{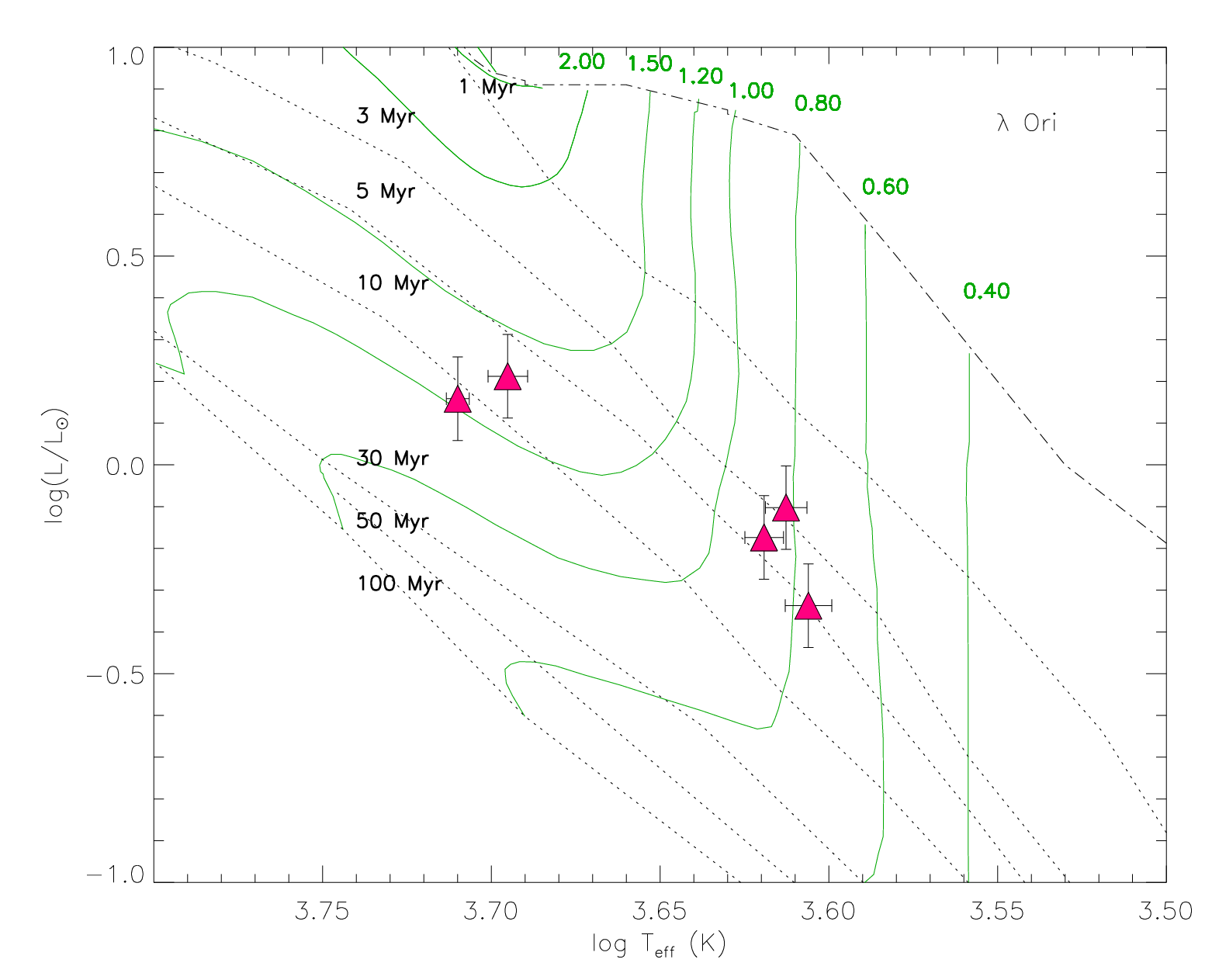}                  
       \caption{HR diagram of the members of 25~Ori ({\it upper panel}; squares) and $\lambda$~Ori ({\it lower panel}; triangles) 
       where the abundance analysis was done. The \cite{pallastahler1999} PMS evolutionary tracks are displayed with 
       the labels representing their masses (solid lines). Similarly, the birthline and isochrones (from 1 to 30 Myr) are shown with 
       dash-dotted and dotted lines, respectively.}
       \label{fig:HRD}
 \end{center}
\end{figure}

As for surface gravity, the initial gravity ($\log g^{\rm in}$ in Table~\ref{tab:param_chem}) was obtained from the relation 
between mass, luminosity, and effective temperature: $\log g=4.44+\log (M/M_\odot) +4 \log (T_{\rm eff}/5770)-\log (L/L_\odot)$. 
Effective temperatures were set to our $T_{\rm eff}^{\rm LDR}$ (see Table~\ref{tab:param_chem}). Stellar luminosities 
(Table~\ref{tab:radvel_lithium}), were computed from a relationship between the absolute bolometric magnitude, the $J$ absolute 
magnitude ($M_{J}$), and the $(V-K)$ color index given by \cite{kenyonhartmann95} for late-type stars, 
considering $M_{\rm bol}^{\odot} = 4.64$ as solar bolometric magnitude (\citealt{cox2000}). The value of $M_{J}$ was obtained adopting
a distance of 330 pc for 25~Ori (\citealt{bricenoetal05}) and 400 pc for $\lambda$~Ori (\citealt{murdinpenston1977}), 
respectively. We used as mean interstellar extinctions the values of $A_V=0.29$ given by \cite{bricenoetal07} for 25~Ori and 
$A_J=0.106$ for $\lambda$~Ori (\citealt{barradoetal2004}). Initial masses ($M^{\rm in}$ in Table~\ref{tab:radvel_lithium}) were 
estimated from the isochrones of \cite{pallastahler1999} at given stellar $M_{J}$ and $T_{\rm eff}^{\rm LDR}$. 
The HR diagrams for both samples are shown in Fig.~\ref{fig:HRD}, together 
with the \cite{pallastahler1999} tracks and isochrones. The 25~Ori stars show an age around 10 Myr, with the exception 
of one member, which is marginally older. Three of the $\lambda$~Ori targets are 3--5 Myr old, while the other 
two are slightly older. Both samples show ages in agreement with previous studies (see \citealt{dolanmathieu2001} 
for $\lambda$~Ori and \citealt{bricenoetal07} for 25~Ori).

\begin{sidewaystable*}
\setlength{\tabcolsep}{1.2pt}
\caption{Astrophysical parameters and elemental abundances derived from spectroscopic analysis.}
\label{tab:param_chem}
\tiny
\begin{center}  
\begin{tabular}{lcccccrrrrrrrrrrrl}
\hline
\hline
Star&$T_{\rm eff}^{\rm LDR}$&$T_{\rm eff}^{\rm SPEC}$& $\tiny{\log g}^{\rm in}$&$\tiny{\log g}^{\rm SPEC}$&$\xi$&[\ion{Fe}{i}/H]&[\ion{Fe}{ii}/H]&[Na/Fe]&[Al/Fe]&[Si/Fe]&[Ca/Fe]&[\ion{Ti}{i}/Fe]&[\ion{Ti}{ii}/Fe]&$<$[Ti/Fe]$>$&[Ni/Fe]\\ 
    & (K) & (K) & & & \tiny{(km/s)} & & & & & & & & & &  &  \\  
\hline
~\\
\multicolumn{16}{c}{25~Ori}\\
\hline
CVSO207& 4447$\pm$48  & 4600 & 4.2& 4.1 & 1.7 &$-$0.10$\pm$0.09(34)& $-$0.07$\pm$0.11(4) &    0.03$\pm$0.10(1)&    0.16$\pm$0.11(2) & 0.06$\pm$0.21(5) &    0.05$\pm$0.09(3) & $-$0.02$\pm$0.16(6)& 0.00$\pm$0.10(2) & $-$0.01$\pm$0.13&$-$0.04$\pm$0.11(13)\\
CVSO211& 4399$\pm$47  & 4550 & 4.3& 4.2 & 1.4 &$-$0.05$\pm$0.09(43)& $-$0.03$\pm$0.13(3) & $-$0.04$\pm$0.09(1)&    0.08$\pm$0.12(2) & 0.08$\pm$0.24(5) &    0.01$\pm$0.10(3) & $-$0.03$\pm$0.12(7)& 0.10$\pm$0.10(2) & 0.04$\pm$0.13   &$-$0.02$\pm$0.12(15) \\  
CVSO214& 4787$\pm$52  & 4900 & 4.3& 4.2 & 2.1 &$-$0.08$\pm$0.06(41)& $-$0.06$\pm$0.09(5) &    0.08$\pm$0.06(1)&    0.12$\pm$0.13(2) & 0.10$\pm$0.10(3) &    0.06$\pm$0.08(3) &    0.01$\pm$0.10(9)& 0.00$\pm$0.08(3) & 0.01$\pm$0.13   &$-$0.02$\pm$0.13(22) \\
CVSO259& 4011$\pm$42  & 4050 & 4.2& ... & 1.4 &   0.01$\pm$0.16(32)&  	    ...          & $-$0.26$\pm$0.16(1)& $-$0.06$\pm$0.22(2) & 0.06$\pm$0.16(2) & $-$0.20$\pm$0.16(2) & $-$0.28$\pm$0.19(8)&      ...	   &        ...      &   0.01$\pm$0.21(6)  \\
CVSO262& 4453$\pm$150 & 4450 & 4.1& ... & 1.9 &$-$0.01$\pm$0.09(23)&  	    ...          &	  ...	      & $-$0.05$\pm$0.09(2) & ...	       & $-$0.04$\pm$0.15(1) & $-$0.21$\pm$0.21(3)&    ...	   &        ...      &$-$0.02$\pm$0.17(2) \\
~\\		   																					       
25~Ori  	 & 	           & &&     &	 &$-$0.05$\pm$0.05 & $-$0.05$\pm$0.02    &    0.02$\pm$0.06   &    0.08$\pm$0.08    & 0.08$\pm$0.02    &    0.02$\pm$0.05    & 	                & 		   & 0.01$\pm$0.02  &$-$0.02$\pm$0.02	 \\		    
\hline
~\\
\multicolumn{16}{c}{$\lambda$~Ori}\\
\hline
DM07   &  4956$\pm$67& 5050 & 4.1& 4.2 & 2.1 &    0.00$\pm$0.07(39)& 0.02$\pm$0.04(3)&   0.06$\pm$0.08(1)&  0.15$\pm$0.01(2)& 0.13$\pm$0.03(4)&$-$0.03$\pm$0.09(2)&   0.00$\pm$0.09(7)&0.12$\pm$0.17(4)& 0.06$\pm$0.20 &$-$0.01$\pm$0.07(14)\\
DM14   &  4037$\pm$64& 4000 & 4.0& ... & 1.4 &    0.01$\pm$0.08(31)&	   ...       &$-$0.26$\pm$0.09(1)& $-$0.13$\pm$0.05(2)& 0.18$\pm$0.19(4)&$-$0.21$\pm$0.09(2)&$-$0.23$\pm$0.12(6)&0.45$\pm$0.14(2)& 0.11$\pm$0.18 &  0.08$\pm$0.13(16)\\
DM21   &  4167$\pm$55& 4300 & 4.0& ... & 1.4 &    0.02$\pm$0.08(19)&	   ...       &$-$0.13$\pm$0.13(1)& $-$0.01$\pm$0.09(2)& 0.06$\pm$0.43(2)&$-$0.09$\pm$0.13(1)&$-$0.05$\pm$0.08(3)&   ...	       &... &$-$0.03$\pm$0.15(3)\\
DM26   &  5128$\pm$40& 5300 & 4.2& ... & 2.3 &    0.01$\pm$0.10(21)&	   ...       &         ...	 &  0.23$\pm$0.01(2)& 0.04$\pm$0.18(3)&        ...	  &   0.06$\pm$0.10(5)&   ...	       & ...&   0.01$\pm$0.13(5)\\
CFHT6  &  4099$\pm$58& 4000 & 3.8& ... & 1.6 & $-$0.01$\pm$0.14(18)&	    ...      &$-$0.23$\pm$0.16(1)& $-$0.20$\pm$0.09(2)& 0.14$\pm$0.17(1)&$-$0.26$\pm$0.14(2)&$-$0.29$\pm$0.20(3)&0.26$\pm$0.19(1)& $-$0.02$\pm$0.28   &   0.04$\pm$0.17(5)\\
~\\
$\lambda$~Ori &       &     &  &&    &    0.01$\pm$0.01   & 0.02$\pm$0.04   &   0.06$\pm$0.08   &  0.19$\pm$0.06   & 0.11$\pm$0.06   & $-$0.03$\pm$0.09  &		      & 	       & 0.05$\pm$0.06  &0.02$\pm$0.04		\\
\hline
\end{tabular}
\end{center}
\end{sidewaystable*}
\normalsize

\subsection{Stellar abundance measurements}
\label{sec:abundance}
Stellar abundances were measured following the prescriptions given by \cite{biazzoetal2011}. We briefly list 
the main steps, but we refer to that paper for a detailed description of the method.
\begin{itemize}
\item We performed abundance analysis in local thermodynamic equilibrium (LTE) conditions using equivalent widths 
and the 2002 version of MOOG (\citealt{sneden1973}). The radiative and Stark broadening were treated in a standard way, 
while, for collisional broadening, we used the \cite{unsold1955} approximation.
\item \cite{Kuru93} and Brott \& Hauschildt (2010, private comm.) grid of plane-parallel model atmospheres were used for 
warm ($>$4400 K) and cool ($\ltsim$4400 K) stars, respectively.
\item Equivalent widths were measured by direct integration or by Gaussian fit using the IRAF {\sc splot} task.
\item We adopted the line list of \cite{biazzoetal2011}, adding other useful iron lines (see Table~\ref{tab:other_iron_lines}) 
taken from \cite{santosetal2008}.
\item After removing lines with $EW>150$ m\AA, likely saturated and most affected by the treatment of damping, we 
applied a $2\sigma$ clipping rejection.
\item Although we do not expect high levels of veiling in our sample, we searched for any possible excess emission. We 
found that all stars had veiling levels consistent with zero, confirming that this quantity is not an issue
in our abundance measurements.
\item Our study is performed differentially with respect to the Sun. We analyzed the {\sc uves} solar spectrum adopting 
$T_{\rm eff}=5770$ K, $\log g=4.44$, $\xi=1.1$ km s$^{-1}$ (see \citealt{randichetal2006}). We obtained 
$\log n{\rm (\ion{Fe}{i})}=7.52\pm0.02$ and $\log n{\rm (\ion{Fe}{i})}=7.51\pm0.02$ with Kurucz and Brott \& Hauschildt 
model atmospheres, respectively. For the iron solar abundance of each line, see Biazzo et al. (2011, and references therein), 
and Table~\ref{tab:other_iron_lines}. The solar abundances of all the other elements are listed in Table~\ref{tab:radvel_lithium} 
of \cite{biazzoetal2011}, since we used the same line list.
\end{itemize}

\begin{table}
\caption{Wavelength, excitation potential, oscillator strength, equivalent width, and abundance obtained with 
ATLAS (\citealt{Kuru93}) and GAIA (Brott \& Hauschildt 2010, private comm.) models for the iron lines taken 
from \cite{santosetal2008} are listed.}
\label{tab:other_iron_lines}
\begin{center}
\begin{tabular}{ccrccc}  
\hline
\hline
$\lambda$ & $\chi$ & $\log gf$ & $EW$ & $\log n_{\rm ATLAS}$& $\log n_{\rm GAIA}$\\
(\AA) & (eV) & & (m\AA) & &  \\
\hline
5322.05 & 2.28 & $-$2.90 & 62.3 & 7.49 & 7.47   \\
5852.22 & 4.55 & $-$1.19 & 41.2 & 7.50 & 7.48  \\
5855.08 & 4.61 & $-$1.53 & 22.2 & 7.46 & 7.45   \\
6027.06 & 4.08 & $-$1.18 & 63.7 & 7.48 & 7.46  \\
6079.01 & 4.65 & $-$1.01 & 48.3 & 7.54 & 7.52 \\
6089.57 & 5.02 & $-$0.88 & 36.1 & 7.51 & 7.49 \\
6151.62 & 2.18 & $-$3.30 & 51.1 & 7.47 & 7.47 \\
6608.03 & 2.28 & $-$3.96 & 17.8 & 7.46 & 7.46  \\
6646.94 & 2.61 & $-$3.94 & 11.0 & 7.51 & 7.51 \\
6710.32 & 1.48 & $-$4.82 & 17.2 & 7.49 & 7.50  \\
\hline
\end{tabular}
\end{center}
\end{table}

\subsubsection{Stellar parameters}
\label{sec:stellar_parameters}

Effective temperatures were also determined by imposing the condition that the \ion{Fe}{i} abundance does not depend on the line 
excitation potentials. The initial value of the effective temperature ($T_{\rm eff}^{\rm LDR}$), and the final values 
($T_{\rm eff}^{\rm SPEC}$) are listed in Table~\ref{tab:param_chem} and plotted in Fig.~\ref{fig:teff_ldr_spec}. The temperatures 
obtained with both methods are in good agreement (mean difference of $\sim$100 K), with $T_{\rm eff}^{\rm SPEC}$ higher on average 
than $T_{\rm eff}^{\rm LDR}$.  

The microturbulence velocity $\xi$ was determined by imposing that the \ion{Fe}{i} abundance is independent on the line equivalent 
width. The initial value was set to 1.5 km\,s$^{-1}$, and the final values are listed in Table~\ref{tab:param_chem}. 

The surface gravity was determined by imposing the ionization equilibrium of \ion{Fe}{i}/\ion{Fe}{ii}. With the exception of 
CVSO259 and CVSO262 with few suitable lines in the spectra, we were able to 
constrain the gravity for the stars in 25~Ori ($\log g^{\rm SPEC}$ in Table~\ref{tab:param_chem}). In the case of $\lambda$~Ori, only DM07 shows enough 
\ion{Fe}{ii} lines to allow a spectroscopic measurement of the gravity.

\subsubsection{Errors}
\label{sec:abun_errors}
As described in \cite{biazzoetal2011}, elemental abundances are affected by random (internal) and systematic (external) errors. 

In brief, sources of internal errors include uncertainties in atomic parameters, stellar parameters, and line equivalent widths, 
where the former should cancel out when the analysis is carried out differentially with respect to the Sun, as in our case. 

Errors due to uncertainties in stellar parameters were estimated by assessing errors in $T_{\rm eff}$, $\xi$, and $\log g$, and 
by varying each parameter separately, while keeping the other two unchanged. We found that variations in $T_{\rm eff}$ larger than 
60 K would introduce spurious trends in $\log n{\rm (\ion{Fe}{i})}$ versus $\chi$; variations in $\xi$ larger than 0.2 km s$^{-1}$ 
would result in significant trends of $\log n{\rm (\ion{Fe}{i})}$ versus $EW$, and variations in $\log g$ larger than 0.2 dex would 
lead to differences between $\log n{\rm (\ion{Fe}{i})}$ and $\log n{\rm (\ion{Fe}{ii})}$ larger than 0.05 dex. Errors in abundances 
due to uncertainties in stellar parameters are summarized in Table~\ref{tab:errors} for one of the coolest stars (DM14) and for the 
warmest target (DM07).

Random errors in EW are well represented by the standard deviation around the mean abundance determined from all the lines. When only 
one line was measured, the abundance error is the standard deviation of three independent EW measurements. In Table~\ref{tab:param_chem} 
these errors are listed together with the number of lines used (in brackets).

As for the external error, the biggest one is given by the abundance scale, which is mainly influenced by errors in 
model atmospheres. As pointed out by \cite{biazzoetal2011}, this error source is negligible, because we used for both 
clusters the same procedure and instrument set-up to measure the elemental abundances.

\begin{table*}  
\caption{Internal errors in abundance determination due to uncertainties in stellar parameters for one of the coolest 
star (namely, DM14) and for the warmest star (DM07). Numbers refer to the differences between 
the abundances obtained with and without the uncertainties in stellar parameters.}
\label{tab:errors}
\begin{center}
\begin{tabular}{lccc}
\hline
\hline
DM14 & $T_{\rm eff}$=4000 K & $\log g=4.0$ & $\xi=1.4$ km/s\\
\hline
$\Delta$   & $\Delta T_{\rm eff}=-/+60$ K & $\Delta \log g=-/+0.2$ & $\Delta \xi=-/+0.2$ km/s\\
\hline
$[$\ion{Fe}{i}/H$]$  & 0.05/$-$0.03 & $-$0.04/0.06 & 0.05/$-$0.04  \\
$[$Na/Fe$]$ & $-$0.09/0.06 & 0.07/$-$0.09 & $-$0.03/0.01	     \\
$[$Al/Fe$]$ & $-$0.05/0.03 & 0.02/$-$0.03  & $-$0.03/0.01  \\
$[$Si/Fe$]$ & 0.04/$-$0.06 & $-$0.05/0.04 & $-$0.05/0.03  \\
$[$Ca/Fe$]$ & $-$0.08/0.07 & 0.07/$-$0.07 &0.01/$-$0.01	 \\
$[$\ion{Ti}{i}/Fe$]$ & $-$0.07/0.06 & 0.03/$-$0.04 & 0.04/$-$0.06  \\
$[$\ion{Ti}{ii}/Fe$]$ & 0.02/$-$0.03 & $-$0.08/0.06 & $-$0.03/0.02  \\
$[$Ni/Fe$]$ & 0.00/$-$0.01 & $-$0.02/0.02 & $-$0.01/0.01  \\				   
\hline	\\
DM07 & $T_{\rm eff}$=5050 K & $\log g=4.2$ & $\xi=2.1$ km/s\\
\hline
$\Delta$ & $\Delta T_{\rm eff}=-/+60$ K & $\Delta \log g=-/+0.2$ & $\Delta \xi=-/+0.2$ km/s\\
\hline
$[$\ion{Fe}{i}/H$]$ &$-$0.01/0.02  & $-$0.02/0.02 & 0.06/$-$0.05  \\
$[$\ion{Fe}{ii}/H$]$& 0.05/$-$0.04 & $-$0.14/0.10 & 0.03/$-$0.02  \\
$[$Na/Fe$]$ & $-$0.03/0.01 & 0.03/$-$0.05 &         $-$0.05/0.03\\
$[$Al/Fe$]$ & $-$0.02/0.01 & 0.03/$-$0.03 &         $-$0.03/0.03\\
$[$Si/Fe$]$ & 0.02/$-$0.04 & $-$0.04/0.00 &         $-$0.05/0.03\\
$[$Ca/Fe$]$ & $-$0.03/0.02 &  0.04/$-$0.04&         $-$0.02/0.01\\
$[$\ion{Ti}{i}/Fe$]$ & $-$0.06/0.04 & 0.02/$-$0.02 & 0.01/$-$0.01\\
$[$\ion{Ti}{ii}/Fe$]$ & 0.03/$-$0.03 & $-$0.08/0.06 & $-$0.05/0.04\\
$[$Ni/Fe$]$ & 0.01/$-$0.02 & $-$0.03/0.01 &           $-$0.03/0.02\\
\hline	\\
\end{tabular}
\end{center}
\end{table*}
 
\subsection{Elemental abundances}
We measured the abundances of iron, sodium, aluminum, silicon, calcium, titanium, and nickel. Our final abundances are listed 
in Table~\ref{tab:param_chem}. 

\begin{figure*}	
\begin{center}
 \begin{tabular}{c}
  \resizebox{\hsize}{!}{\includegraphics{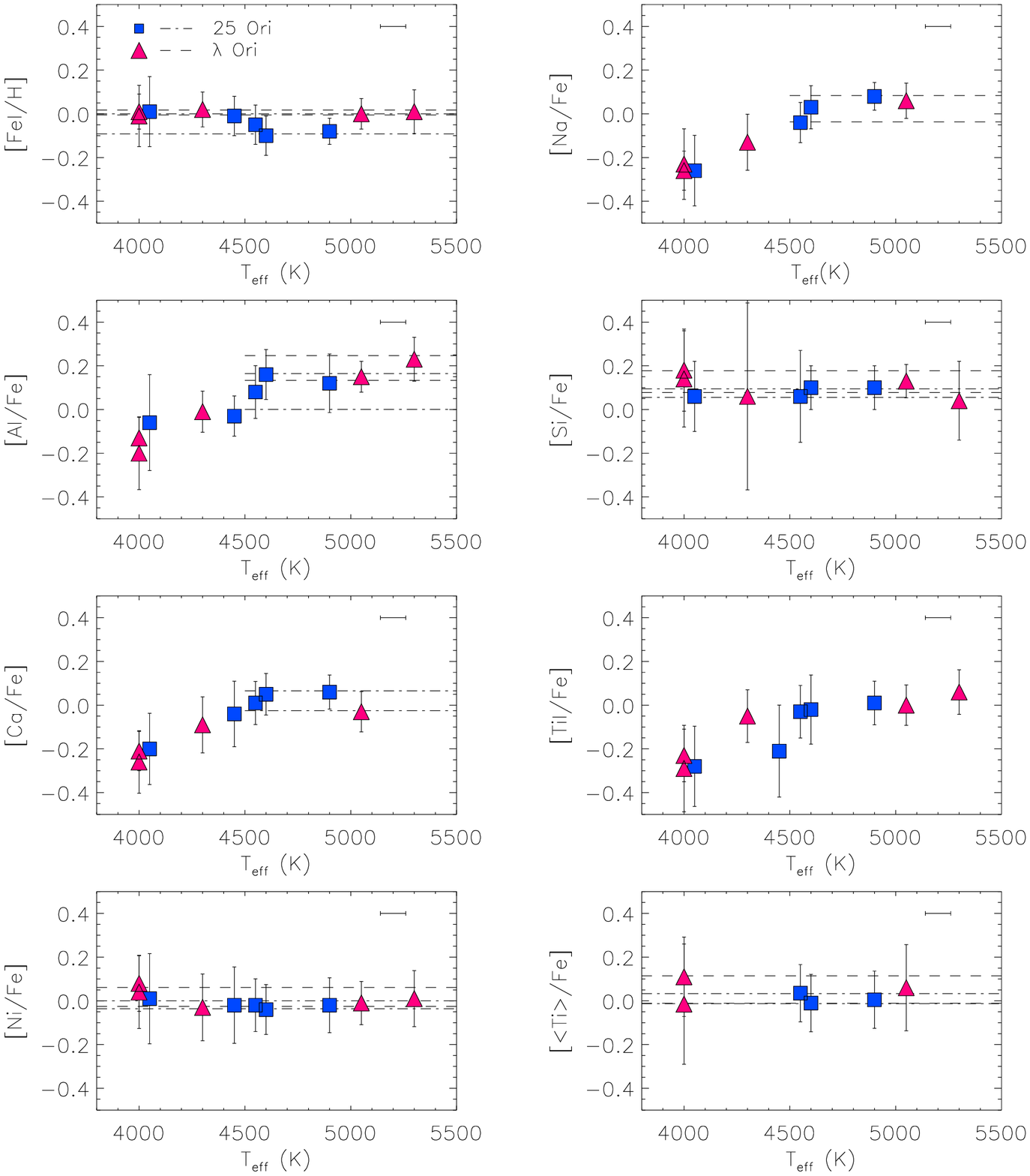}}\\                  
 \end{tabular}
 \vspace{-1.5cm}
       \caption{[X/Fe] versus $T_{\rm eff}^{\rm SPEC}$ for our 25~Ori (squares) and $\lambda$~Ori (triangles) members. 
       Mean abundances values for elements affected by NLTE effects were computed considering the stars with 
       $T_{\rm eff}^{\rm SPEC} > 4400$ K, while $\pm1\sigma$ dispersions are shown by dash-dotted lines for 25~Ori 
       and dotted lines for $\lambda$~Ori. The dashed and dash-dotted lines show the $\pm1\sigma$ values from 
       the mean 25~Ori and $\lambda$~Ori abundances, respectively. The horizontal error bar in all plots represents the 
       typical uncertainty in $T_{\rm eff}^{\rm SPEC}$.}
       \label{fig:elements_abund}
 \end{center}
\end{figure*}

In Fig.~\ref{fig:elements_abund} we show the [Fe/H] and [X/Fe] ratios as a function of $T_{\rm eff}^{\rm SPEC}$ for 
25~Ori and $\lambda$~Ori. The figure shows that there is no star-to-star variation in all elements 
in both groups with the exceptions of the elements where NLTE effects are present, namely Na, Al, Ca, 
and Ti. In particular, due to their young age, cluster stars are characterized by high levels of chromospheric 
activity and are more affected by NLTE over-ionization, which cause a decreasing trend towards lower temperatures of 
neutral species of elements with low ionization potential (see \citealt{dorazirandich09, biazzoetal2011}, and 
references therein, for thorough discussions of this issue).

For 25~Ori, the mean iron abundance is $<$[Fe/H]$>=-0.05\pm 0.05$, while for $\lambda$~Ori we find $<$[Fe/H]$>$$=0.01\pm 0.01$. 

Fig.~\ref{fig:elements_abund} also shows that stars of 25~Ori and $\lambda$~Ori are characterized by homogeneous, close-to-solar 
elemental abundances.

\section{Discussion}
\label{sec:discussion}
\subsection{Metallicity distribution in Orion}
\label{sec:met_distr}
\begin{figure}	
\begin{center}
 \includegraphics[width=9cm]{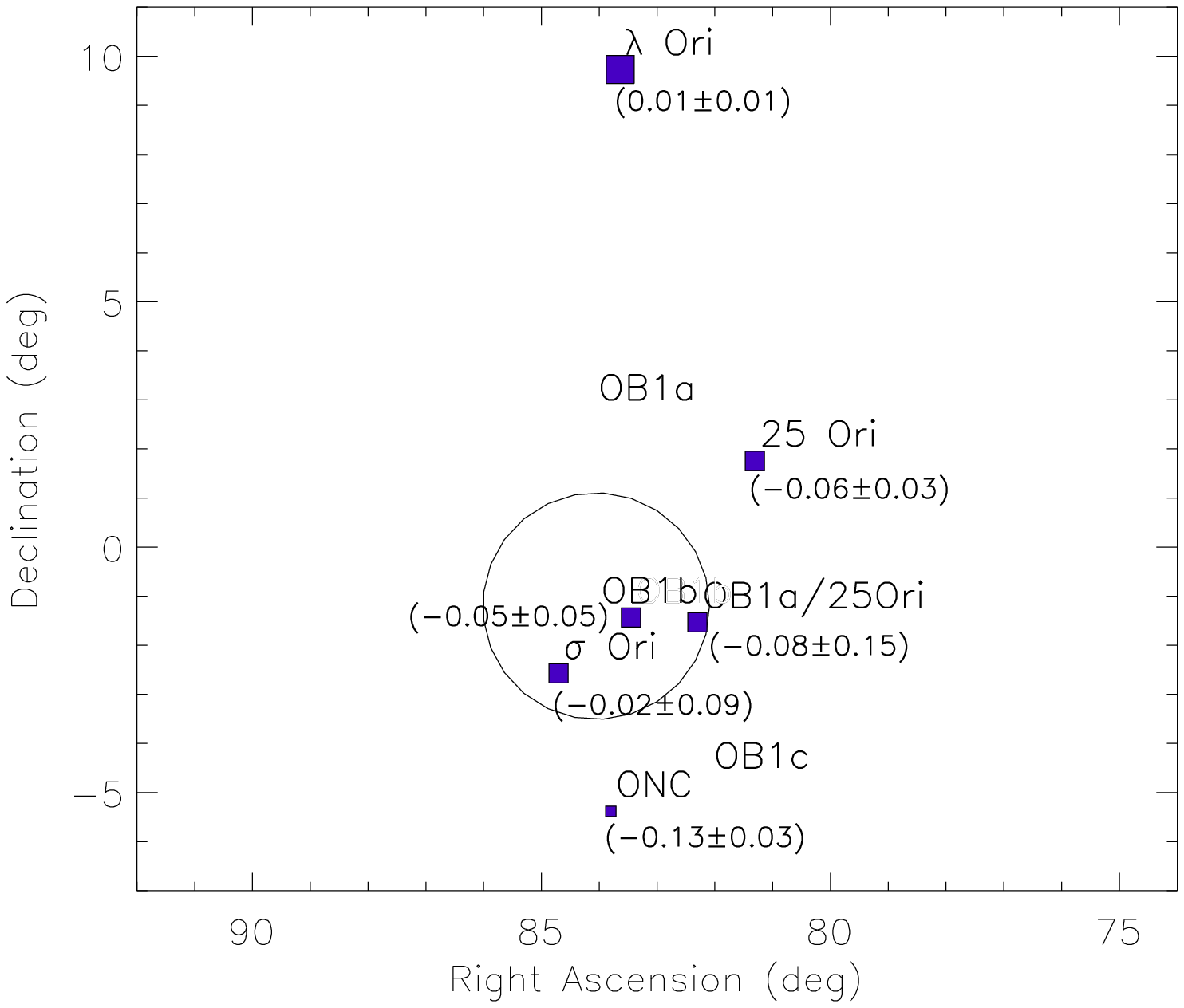}
       \caption{Iron abundance distribution of members of the Orion complex (ONC, OB1b, OB1a, $\sigma$~Ori, 
       25~Ori, and $\lambda$~Ori) taken from the most recent literature (\citealt{gonzalez-hernandez2008, biazzoetal2011}) 
       and this work. Symbols with different size denote three different metallicity bins: [Fe/H]$\le-0.10$, 
       $-0.09\le$[Fe/H]$\le0.00$, and [Fe/H]$> 0.00$, from the smallest to the largest size. The central bin 
       corresponds to the average metallicity of the Orion complex $\pm 1\sigma$. The mean metallicity of each 
       subgroup is written in parenthesis. The circle outlines the \cite{bricenoetal05} boundary of the 
       Orion OB subgroups.}
       \label{fig:orion_iron_distr}
 \end{center}
\end{figure}

In Fig.~\ref{fig:orion_iron_distr} {and in Table~\ref{tab:sfr_iron_abun}} we show the [Fe/H] distribution of the Orion subgroups, 
namely the Orion Nebula Cluster (ONC), OB1b, OB1a (\citealt{biazzoetal2011}), $\sigma$~Ori 
(\citealt{gonzalez-hernandez2008}), $\lambda$~Ori, and 25~Ori (this work). As can be seen, the Orion complex seems to be 
slightly inhomogeneous, with the iron abundance ranging from $-$0.13$\pm$0.03 (the ONC) to 0.01$\pm$0.01 ($\lambda$~Ori), and 
an internal dispersion of $\sim 0.05$ dex. Interestingly, while $\lambda$~Ori has an iron abundance identical to solar, 
all the Orion subgroups are below solar. It is also remarkable that the youngest object, the ONC, has the lowest 
value at all. These results seem to exclude a direct and significant amount of contamination between neighboring (or adjacent) 
regions, or an increase of iron abundance with age, as na\"{\i}vely expected in a SN-driven abundance enriched scenario 
(see \citealt{biazzoetal2011}, and references therein). This is also supported by the velocity dispersion observed in the 
Orion subgroups, which is around 0.5--2.7 km/s, or $\sim$0.5--2 pc/Myr (see Table~\ref{tab:sfr_iron_abun}). This means 
that the members of the subgroups would have moved at most (if on ballistic orbits) by only a few pc for the ONC 
and up to $\sim$10 pc for the oldest ones. Excluding systematic biases in the analysis, this small dispersion in Orion may be 
the result of several effects. $i)$ Different and independent episodes of star formation between $\lambda$~Ori and 
the other Orion subgroups analyzed. The former formed $\sim 8-10$ Myr ago from a supernova explosion close to the 
present $\lambda^1$~Ori position (\citealt{dolanmathieu1999, dolanmathieu2001}), while the other regions are the result of 
sequential star formation triggered by supernova type-II events in the OB1a association (\citealt{preibzinne2006}). 
$ii)$ Large-scale formation processes on $\sim 1$ kpc scale may lead to chemically inhomogenous gas inside the giant 
molecular cloud (\citealt{elmegreen1998}). A quick onset of star formation in clumps of the molecular clouds born under 
turbulent conditions implies that any non-uniformity of abundances may enter the cloud, from an initial non-uniform 
pre-cloud gas or from the background Galactic gradient. In this case there is not enough time for uniform mixing before star 
formation, and the range of cloud metallicity inhomogeneity may reach a level of $\pm 0.05$ dex. Even if the spatial scales 
involved in the formation of the Orion complex are most likely smaller (\citealt{hartmannburkert2007}), the presence 
of inhomogeneous gas in the interstellar medium could contribute to the observed small dispersions. 

These results are encouraging, because for many years our knowledge about the chemical composition in the Orion complex 
from early-B main sequence stars (\citealt{cunhalamb92, cunhalamb94}) and late-type stars (\citealt{cunhaetal98, dorazietal09}) 
was characterized by highly inhomogeneous abundances, with group-to-group differences of $\sim 0.40$ 
dex in oxygen, $\sim 0.30-0.40$ dex in silicon, and $\sim 0.10$ dex in iron. These large spreads have been interpreted as 
the chemical signature of self-enrichment inside the Orion OB association by ejecta of type-II SNe from massive 
stars. Our results change this picture. Also, the new revision of the abundances of B-type stars in Orion 
OB1a, b, c, d performed by \cite{simondiaz2010} and the good agreement between our own and the \cite{santosetal2008} 
values for low-mass stars indicate a very low dispersion in the abundances of oxygen, silicon, nickel, and iron, 
with the Orion Nebula Cluster representing the most metal-poor subgroup. Note that, as pointed out by \cite{simondiaz2010} 
and \cite{biazzoetal2011}, the small group-to-group elemental 
abundance dispersions are smaller than the internal errors in measurements.

Why is the ONC so significantly different from the other regions is an interesting question that should be addressed 
in the future. The cluster is located at the tip of a loose molecular cloud that contains an extended population 
of embedded and optically visible low-mass stars (\citealt{megeathetal2005}). Their properties have been recently 
analyzed by \cite{fangetal2009}, who provide useful candidates to be observed for abundance measurements. Then, one 
can verify if the ONC is just an anomaly or if it shows the same abundance of the present Giant Molecular Cloud. Similarly, 
the situation in $\lambda$~Ori is favorable to study a possible link between what we find close to the $\lambda^1$~Ori 
star and the chemical abundances of the current generation of young stars located in the vicinity of the molecular 
gas clumps shown in Fig.~\ref{fig:lambdaOri_map}. For example, \cite{dolanmathieu2001} provide the properties 
of the CTTS/WTTS in these regions that can then be observed for metallicity measurements.

\begin{table*}
\caption{Mean iron abundance of low-mass stars, velocity dispersion, and age range of the Orion subgroups.}
\label{tab:sfr_iron_abun}
\tiny
\begin{center}
\begin{tabular}{crlr|cl|cl}  
\hline
\hline
SFR & [Fe/H] & Reference & \# stars & $\delta V_{\rm rad}$ & Reference & Age & Reference\\
& &  & & (km/s) & &  (Myr) & \\
\hline
$\sigma$~Ori     & $-$0.02$\pm$0.09 & \cite{gonzalez-hernandez2008}  & 8  & 0.9     & \cite{saccoetal2008} & 2--3 & \cite{walteretal2008}\\
ONC              & $-$0.13$\pm$0.03 & \cite{biazzoetal2011}          & 10 & 3.1     & \cite{fureszetal2008}& 2--3 & \cite{darioetal2010}\\
OB1b             & $-$0.05$\pm$0.05 & \cite{biazzoetal2011}          & 4  & 1.9     & \cite{bricenoetal07} & 4--6 & \cite{bricenoetal07}\\
OB1a/25~Ori      & $-$0.08$\pm$0.15 & \cite{biazzoetal2011}          & 1  & $\sim$2 & \cite{bricenoetal07} & 7--10& \cite{bricenoetal05} \\
25~Ori           & $-$0.05$\pm$0.05 & This work                      & 5  & 1.7     & \cite{bricenoetal07} & 7--10 & \cite{bricenoetal05}\\
$\lambda$~Ori    &    0.01$\pm$0.01 & This work                      & 5  & 0.5     & \cite{saccoetal2008} & 5--10 & \cite{dolanmathieu2002}\\
\hline
\end{tabular}
\end{center}
\end{table*}

\subsection{SFR abundances in the Galactic disk}

Each component of our Galaxy (bulge, halo, thin and thick disk) presents a characteristic abundance pattern of elements (such 
as iron-peak and $\alpha$-elements) with respect to iron. The observed differences reflect a variety of star formation 
histories. One would expect that SFRs show a similar pattern as thin disk stars, unless the gas from which they have 
formed has undergone a peculiar enrichment. In order to check if this is indeed true, we compare [X/Fe] ratios versus 
[Fe/H] trends for SFRs (\citealt{kingetal2000, santosetal2008, gonzalez-hernandez2008, biazzoetal2011, dorazietal2011}) 
with studies of nearby field stars of the thin Galactic disk (\citealt{soubirangirard2005}). 

\begin{figure*}	
\begin{center}
 \begin{tabular}{c}
 \includegraphics[width=16cm]{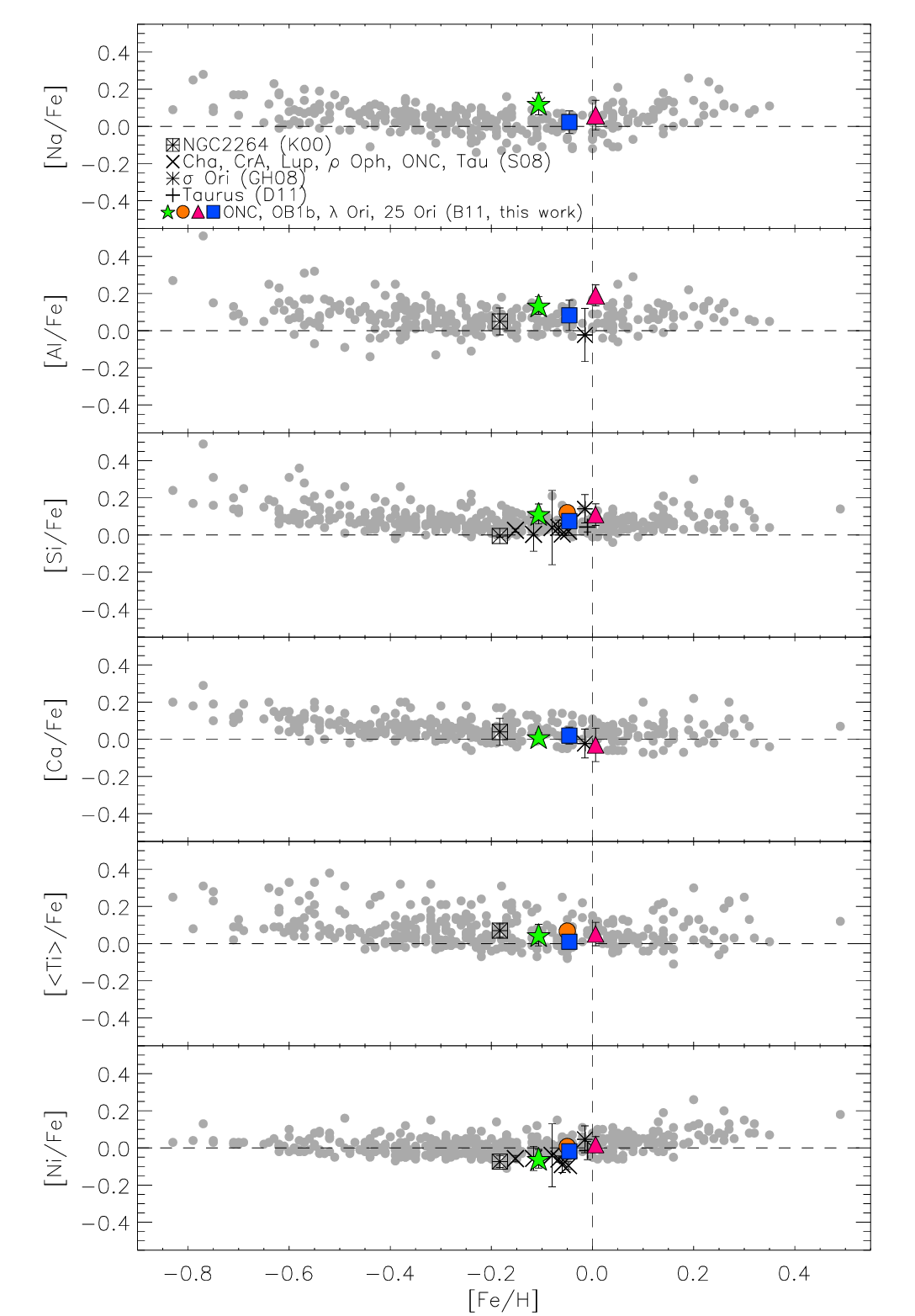}
 \end{tabular}
       \caption{[X/Fe] versus [Fe/H] for SFRs with known high-resolution abundances. The references are: Biazzo et al. (2011; B11), 
       D'Orazi et al. (2011; D11), Santos et al. (2008; S08), King et al. (2000; K00), and Gonz\'alez-Hern\'andez et al. (2008; 
       GH08). The errors in [Fe/H] are around 0.02--0.12 dex. The dashed lines mark the 
       solar abundances. The filled dots in the background represent the [X/Fe] distribution of nearby field stars in the 
       Galactic thin disk (\citealt{soubirangirard2005}).}
       \label{fig:alphaFe_FeH}
 \end{center}
\end{figure*}

Fig.~\ref{fig:alphaFe_FeH} shows an overall good agreement between the [X/Fe] versus [Fe/H] distribution of SFRs and 
nearby field stars. A more detailed discussion on the different groups of elements is provided below:

\paragraph{The $\alpha$ elements Si, Ca, and Ti}
~\\
The $\alpha$ elements are mostly produced in the aftermath of explosions of type II supernovae, with small contribution from 
type Ia SNe. On average, [Si/Fe] in SFRs seems to be slightly higher than the Sun (Fig.~\ref{fig:alphaFe_FeH}). Given the 
dispersion in Si abundance of the SFRs, it is fully consistent with the distribution of the field stars of the thin disk. 
The behaviour of [Ca/Fe] and [Ti/Fe] ratios versus [Fe/H] is similar to those reported by \cite{soubirangirard2005} for 
field stars with $-0.2\ltsim$[Fe/H]$\ltsim0.2$.

The slight overabundance in [$\alpha$/Fe] of the SFRs might be the result of the contributions of recent local enrichment 
due to SNIIe ejecta (occurred in particular in the Orion complex) or of local accretion from enriched gas pre-existing in 
the nearby thin disk (\citealt{elmegreen1998}). In particular, the slight overabundance in Si of SFRs might reflect 
its larger production during SNII events compared to the calcium and titanium elements (\citealt{woosleyweaver1995}).

\paragraph{The iron-peak element nickel}
~\\
Most of the iron-peak elements are synthesized by SNIa explosions. The nickel abundance is very close to the solar value, 
as found for field stars with [Fe/H] between $-0.4$ and 0.0 (Fig.~\ref{fig:alphaFe_FeH}).

\paragraph{Other elements: Na and Al}
~\\
Sodium and aluminum are thought to be mostly a product of Ne and Mg burning in massive stars, through the NeNa and MgAl 
chains. The [Na/Fe] ratio does not show a trend with [Fe/H], with a mean value close to that of field stars at the same [Fe/H]. 
Similarly, the Al abundance does not show trend with [Fe/H], but at slightly above-solar values, as for field stars 
in the solar neighborhood. Its phenomenological behaviour is very similar to the $\alpha$-elements, as indicated by 
\cite{McWilliam1997}.

\section{Conclusions}
\label{sec:conclusion}
In this paper we have derived homogeneous and accurate abundances of Fe, Na, Al, Si, Ti, Ca, and Ni of two 
very young clusters, namely 25~Orionis and $\lambda$~Orionis. Our main results can be summarized as follows: 

\begin{itemize}
\item Stellar properties have been derived for 14 low-mass stars over a relatively wide temperature interval. 
Four new PMS candidates show kinematics consistent with membership in the 25~Ori cluster.
\item $\lambda$~Ori and 25~Ori have mean iron abundances of $0.01\pm0.01$ and $-0.05\pm0.05$, respectively, without the presence 
of metal-rich stars.
\item Whereas this should be confirmed based on larger number statistics, our results suggest that no 
star-to-star variation is present for all elements, with a high degree of homogeneity inside each stellar group. 
\item Elemental abundances of $\lambda$~Ori, 25~Ori, and other known SFRs agree with those of the Galactic thin disk.
\item The Orion complex seems to be characterized by a small group-to-group dispersion, consistent with 
a different star formation history of each subgroup and an initially inhomogeneous interstellar gas.
\item Finally, we note that the close-to-solar metallicity of both 25~Ori and $\lambda$~Ori reinforces the conclusion that none of 
the SFRs in the solar neighborhood is metal-rich and that metal-rich stars hosting giant planets are likely migrated from the 
inner part of the Galactic disk to their current location (\citealt{santosetal2008, haywood2008}).
\end{itemize}

\begin{acknowledgements}
The authors are very grateful to the referee for a careful reading of the paper. 
KB thanks the INAF--Arcetri Astrophysical Observatory for financial support. Thomas Dame kindly provided the CO maps. This research has made 
use of the SIMBAD database, operated at CDS (Strasbourg, France). 
\end{acknowledgements}

\bibliographystyle{aa}

\end{document}